\documentclass{aa}  
\usepackage{graphicx}
\usepackage{txfonts}
\usepackage{lscape}
\usepackage{placeins}

\def\hi{\relax \ifmmode {\mbox H\,\texjtsc{i}}\else H\,{\scshape i}\fi}
\def\hii{\relax \ifmmode {\mbox H\,\textsc{ii}}\else H\,{\scshape ii}\fi}
\def\nii{\relax \ifmmode {\mbox N\,\textsc{ii}}\else N\,{\scshape ii}\fi}
\def\oi{\relax \ifmmode {\mbox O\,\textsc{i}}\else O\,{\scshape i}\fi}
\def\oii{\relax \ifmmode {\mbox O\,\textsc{ii}}\else O\,{\scshape ii}\fi}
\def\oiii{\relax \ifmmode {\mbox O\,\textsc{iii}}\else O\,{\scshape iii}\fi}
\def\sii{\relax \ifmmode {\mbox S\,\textsc{ii}}\else S\,{\scshape ii}\fi}

\begin{document}

\title{Stellar mass is not the best predictor of galaxy metallicity}
\subtitle{The gravitational potential-metallicity relation $\Phi \rm ZR$}
\titlerunning{The gravitational potential-metallicity relation $\Phi \rm ZR$}

\author{Laura S\'anchez-Menguiano\inst{1,2}
\and
Jorge S\'anchez Almeida\inst{3,4}
\and
Sebasti\'an F. S\'anchez\inst{5}
\and
Casiana Mu\~noz-Tu\~n\'on\inst{3,4}
          }
\authorrunning{L. S\'anchez-Menguiano et al.}

\institute{Dpto. de F\'isica Te\'orica y del Cosmos, Facultad de Ciencias (Edificio Mecenas), Universidad de Granada, E-18071 Granada, Spain
\and
             Instituto Carlos I de F\'isica Te\'orica y computacional, Universidad de Granada, E-18071 Granada, Spain\\
              \email{lsanchezm@ugr.es}
\and
            Instituto de Astrof\'isica de Canarias, E-38205 La Laguna, Tenerife, Spain
\and
            Universidad de La Laguna, Dpto. Astrof\'isica, E-38206 La Laguna, Tenerife, Spain
\and
             Universidad Nacional Aut\'onoma de M\'exico, Instituto de Astronom\'ia, AP 106, Ensenada 22800, BC, M\'exico
             }

 \date{Received --- / Accepted ---}

 
  \abstract
{Interpreting the scaling relations followed by galaxies is a fundamental tool for assessing how well we understand galaxy formation and evolution. Several scaling relations involving the galaxy metallicity have been discovered through the years, the foremost of which is the scaling with stellar mass. This so-called mass-metallicity relation is thought to be fundamental and has been subject to many studies in the literature.} 
{We study the dependence of the gas-phase metallicity on many different galaxy properties to assess which of them determines the metallicity of a galaxy.}
{We applied a random forest regressor algorithm on a sample of more than 3000 nearby galaxies from the SDSS-IV MaNGA survey. Using this machine-learning technique, we explored the effect of 148 parameters on the global oxygen abundance as an indicator of the gas metallicity.}
{$M_{\rm \star}$/$R_e$, as a proxy for the baryonic gravitational potential of the galaxy, is found to be the primary factor determining the average gas-phase metallicity of the galaxy ($Z_g$). It outweighs stellar mass. A subsequent analysis provides the strongest dependence of $Z_g$ on $M_\star / R_e^{\,0.6}$. We argue that this parameter traces the total gravitational potential, and the exponent $\alpha\simeq 0.6$ accounts for the inclusion of the dark matter component.}
{Our results reveal the importance of the relation between the total gravitational potential of the galaxy and the gas metallicity. This relation is tighter and likely more primordial than the widely known mass-metallicity relation.
}


   \maketitle
%

\section{Introduction}

The gas-phase metallicity ($Z_g$) of local galaxies has proven to be a key observable for numerical simulations and theoretical models. As a consequence of recurrent star formation processes, the amount of metals in galaxies increases gradually. In this way, every internal and external process that a galaxy experiences leaves imprints on the chemical distribution of its gas. This parameter has been widely confirmed as a powerful tool for improving our knowledge of the formation and evolution of galaxies \citep[][among many others]{mosconi2001, lia2002, brooks2007, oppenheimer2008, dave2011, pilkington2012, molla2019, spitoni2019}.

Many works have reported significant correlations both globally and locally between the gas metallicity and other galaxy properties, such as the stellar mass \citep[$M_\star$,][]{lequeux1979, tremonti2004, sanchez2017}, the star formation rate \citep[SFR,][]{mannucci2010, sanchezmenguiano2019, sanchezalmeida2019}, the stellar age \citep{lian2015, sanchezmenguiano2020b}, the rotation velocity \citep{garnett2002, pilyugin2004}, the gas mass fraction \citep{bothwell2013, barreraballesteros2018}, and the galaxy size \citep{ellison2008, sanchezalmeida2018b}. Of all galaxy properties, $M_\star$ has been shown to present the strongest correlation with $Z_g$, following the so-called mass-metallicity relation (MZR). As the stellar mass increases, the metallicity increases, until the relation flattens at high masses \citep[e.g.][]{tremonti2004, kewley2008, sanchez2013, wu2016,  barreraballesteros2017, sanchez2017, sanchez2019b, curti2020}. The MZR is found to exist up to $z\sim3.5$ with a similar shape, although it evolves with redshift such that high-z galaxies are less enriched than local ones for a given stellar mass, especially at low masses \citep{maiolino2008, mannucci2009, yabe2012, maier2014, zahid2014, sanders2015, hidalgo2017}. This relation has been proposed to be the result of the interplay of many processes such as metal removal by outflows, dilution by metal-poor gas infall, enrichment by previous episodes of star formation, or simple evolution by the so-called downsizing, where most massive galaxies exhaust their gas reservoir faster \citep{maiolino2019}.

Despite the relevance confirmed for the MZR in the literature, a recent study by \citet{deugenio2018} suggested that gas metallicity correlates more tightly with the average gravitational potential (estimated as $\Phi = M_{\star}/R_{e}$, where $R_e$ is the effective radius of the galaxy) than with either the stellar mass or the average mass surface density ($\Sigma_{\star} \equiv M_{\star}/R_{e}^2$). Using a volume-limited sample drawn from the publicly available Sloan Digital Sky Survey Data Release 7 (SDSS DR7), their conclusions were based on (i) a lower scatter in the relation, (ii) a higher Spearman rank correlation coefficient, and (iii) a weaker residual trend with $R_e$. The three relations were explained in terms of metal enrichment (driven by the gas fraction, and therefore $\Sigma_{\star}$) and metal loss (proportional to the stellar mass). Alternatively, they could emerge as a consequence of an existing local relation between $Z_g$ and the escape velocity (e.g., reported in \citealt{barreraballesteros2018}), a local quantity directly related with the global gravitational potential. In this way, the potential–metallicity relation would be the primary relation, and the MZR and the $\Sigma$ZR would arise from the local relation between $Z_g$ and $\Phi$. This result contradicts another recent study by \citet{baker2023c}, who find a tighter relation between $Z_g$ and $M_{\star}$ than between $Z_g$ and the dynamical mass ($M_{\rm dyn}$) or $Z_g$ and the gravitational potential (estimated as $\Phi = M_{\rm dyn}/R_{e}$). We note, however, that dynamical masses were derived via Jeans anisotropic modelling, which is a complex method that might increase the final scatter and thus undermine its importance.

In general, studies focus on investigating particular relations between the gas metallicity and other galaxy properties, but we lack works that address the issue in a holistic approach, involving a large number of parameters in the analysis. This is key for assessing with accuracy which property predicts $Z_g$ best, that is, which property yields the tightest relation with $Z_g$, and so more probably is the primary relation. In this regard, the recent study by \citet{alvarezhurtado2022} explored the impact of 29 physical parameters on the gas metallicity for a sample of 299 star-forming galaxies from the CALIFA survey \citep[Calar Alto Legacy Integral Field Area,][]{sanchez2012a}. The results of the analysis indicate that the stellar mass is the physical parameter that traces the observed $Z_g$ better. However, they did not include the gravitational potential as a parameter in the analysis. 

This paper describes our work to provide a robust answer to the question. In order to do this, we applied a random forest regressor algorithm on a sample of more than 3000 nearby star-forming galaxies from the MaNGA survey \citep[Mapping Nearby Galaxies at Apache Point Observatory,][]{bundy2015}. Using this technique, we explore the effect of 148 parameters on the global oxygen abundance of the galaxies as an indicator of the gas metallicity. We include in the model all properties that have previously been reported to strongly correlate with $Z_g$, such as $M_{\star}$, $\Phi$, $\Sigma_{\star}$, and $M_{\rm gas}$.

The paper is organised as follows. Section~\ref{sec:manga} briefly presents the MaNGA data, and the description of the procedure we followed in the analysis is given in Section~\ref{sec:analysis}, including the physical parameters of the model (Section~\ref{sec:parameters}), the criteria we adopted for the sample selection (Section~\ref{sec:sample}), and an overview of the random forest algorithm (Section~\ref{sec:RF}). The outcome of the random forest is described in Section~\ref{sec:results1}, and a subsequent analysis including different combinations of $M_{\star}$ and $R_e$ is described in Section~\ref{sec:results2}. All the results are discussed in Section~\ref{sec:dis}, and Section~\ref{sec:concl} compiles and summarises the main conclusions of the work. Finally, supplementary material is included in Appendices \ref{sec:appendix1} and \ref{sec:appendix2}. Appendix~\ref{sec:appendix1} presents the complete list of all the galactic parameters included in the model. Appendix~\ref{sec:appendix2} reproduces the analysis using alternative calibrations to estimate $Z_g$.


\section{MaNGA data}\label{sec:manga}

We took advantage of the large statistics provided by the MaNGA survey \citep[][]{bundy2015}, a now complete project based on integral field spectroscopy (IFS) and part of the fourth-generation SDSS (SDSS-IV). With all the data publicly released \citep[DR17,][]{abdurrouf2022}, MaNGA has gathered spatially resolved information for $10\,010$ galaxies up to redshift $\sim0.15$. 

The data were collected using the two BOSS spectrographs mounted on the Sloan 2.5\,m telescope at Apache Point Observatory \citep{gunn2006}. The 17 simultaneously available fibre bundles are distributed in five different hexagonal configurations, whose field of view (FoV) varies from $12.5''$ to $32.5''$ in diameter \citep{drory2015}. The covered wavelength range spans from $3600$ \AA\ to $10300$ \AA, with a nominal resolution of $\lambda/\Delta\lambda \sim 2100$ at 6000 \AA\ \citep{smee2013}. The resulting spectra for each sampled spaxel of $0.5'' \times 0.5''$ present a final spatial resolution of FWHM $\sim2.5''$, which corresponds to a physical resolution of $\sim 1.5$ kpc at an average redshift of 0.03.

Details of the MaNGA mother sample, the survey design, the observational strategy, and the data reduction are provided in \citet{law2015}, \citet{yan2016}, \citet{law2016}, \citet{wake2017}, and \citet{law2021}.

\section{Analysis}\label{sec:analysis}
\subsection{Galaxy parameters}\label{sec:parameters}

In order to determine the galactic property that best predicts the gas metallicity, we explored the effect of 148 parameters. These parameters were extracted from the {\tt pyPipe3D} Value Added catalogue \citep[VAC,][]{sanchez2022}, which is publicly accessible through the SDSS-IV VAC website\footnote{\url{https://www.sdss4.org/dr17/manga/manga-data/manga-pipe3d-value-added-catalog/}}. This VAC contains the dataproducts of {\tt pyPipe3D} for the full DR17 MaNGA sample (\citealt{lacerda2022}, see also \citealt{sanchez2016a, sanchez2016b}). This analysis pipeline was developed to characterise the properties of the stellar populations and the ionised gas. The catalogue comprises a single table with the integrated values, those measured at the effective radius, and the radial gradients of different properties (e.g. stellar mass, star-formation, oxygen and nitrogen abundances, or dust attenuation), as well as a FITS file per galaxy with the corresponding spatially resolved quantities. 

\begin{table*}[h]
\caption{Examples of the analysed parameters from the {\tt pyPipe3D} VAC.}             
\label{tab1}      
\centering          
\begin{tabular}{l@{\hspace{0.8cm}}l@{\hspace{0.8cm}}l}   
\hline\hline\\[-0.3cm]
Name & Unit & Description \\
\hline\\[-0.3cm]
O/H$_{Re}$ & dex & Oxygen abundance at the effective radius ($R_e$) using the M13 O3N2-calibrator\\[0.1cm]
$\Phi_{\rm baryon}$& log(M$_\odot$/kpc) & Baryonic gravitational potential at $R_e$ estimated as $M_{\rm \star, \, phot}$/$R_e$\\[0.1cm]
$M_{\rm \star, \, phot}$ & log(M$_\odot$) & Integrated stellar mass obtained from photometry\\[0.1cm]
LW-[Z/H]$_{Re}$ & dex & Luminosity weighted (LW) stellar metallicity at $R_e$, normalised to the solar value, \\[0.1cm]
& & in logarithmic scale\\[0.1cm]
MW-[Z/H]$_{Re}$ & dex & Mass weighted (MW) stellar metallicity at $R_e$, normalised to the solar value, in \\[0.1cm]
& & logarithmic scale\\[0.1cm]
$g-r$& mag & $g-r$ colour extracted from the NSA catalogue\\[0.1cm]
T99-[Z/H]$_{Re}$ & dex & Stellar metallicity at $R_e$, normalised to the solar value, in logarithmic scale at T99\\[0.1cm]
& & (look-back time at which the galaxy has formed 99\% of its mass)\\[0.1cm]
$B-R$& mag & $B-R$ colour computed from the MaNGA data\\[0.1cm]
T95-[Z/H]$_{Re}$ & dex & Stellar metallicity at $R_e$, normalised to the solar value, in logarithmic scale at T95 \\[0.1cm]
& & (look-back time at which the galaxy has formed 95\% of its mass)\\[0.1cm]
$B-V$& mag & $B-V$ colour computed from the MaNGA data\\[0.1cm]
$A_{V, \,Re}$& mag & Dust attenuation in the V-band derived from the H$\alpha$/H$\beta$ line ratio at $R_e$\\[0.1cm]
\hline                  
\end{tabular}
\tablefoot{The table only displays the most relevant analysed parameters (see Sec.\ref{sec:results1}, Figure~\ref{fig2}) from the {\tt pyPipe3D} VAC. The complete list is provided in Appendix~\ref{sec:appendix1}. We refer to \citet{sanchez2022} for more details of the derivation of the parameters.}
\end{table*}

For this study, we focused on the global galaxy properties provided in the VAC table. This table comprises 304 parameters (and in most cases, their associated error estimates), many of which correspond to different line intensity values or ratios (e.g. the logarithm of the [NII]6583/H$\alpha$ line ratio in the central $2.5\arcsec$ aperture). These quantities were discarded from the study because including them would result in obvious dependences of some galaxy properties and the emission lines from which they are derived (e.g. between the star formation rate and the H$\alpha$ intensity). In addition, as a proxy for the global gas metallicity ($Z_g$), we used the oxygen abundance (O/H) measured at one disc effective radius ($R_e$)\footnote{This quantity is derived performing a linear fit to the radial profile between 0.5 and 1.5 $R_e$ and evaluating it at 1 $R_e$, as described in \citet{sanchez2022}. The correlation between these values and the average metallicities at 1 $R_e$ --measured in an elliptical ring of width 0.15 $R_e$-- is very high ($r=0.994$). This ensures that the $Z_g$ values used in the analysis are indeed representative of the global galaxy metallicities.} and determined from the empirical calibration proposed by \citet[][hereafter M13]{marino2013} for the O3N2 index. Other $Z_g$ calibrators were employed and led to results similar to those described in Sec.~\ref{sec:results1} for O3N2 (see Appendix~\ref{sec:appendix2} for a deeper discussion). All these additional measurements of O/H were also discarded from the main analysis for clear reasons. Finally, the ionisation parameter, the nitrogen abundance, and the N/O abundance were also ignored because their determination involves line ratios in common with the O/H derivation. All in all, the final selection of galaxy properties available for the analysis ($N_{\rm par}$) comprises 148 parameters. Table~\ref{tab1} compiles the most representative galactic properties, and the complete list is detailed in Appendix~\ref{sec:appendix1}.

\subsection{Sample selection}\label{sec:sample}

The galaxy sample we analysed was drawn from the 10\,010 galaxies comprising the MaNGA mother sample. In order to select the galaxies, we adopted two simple criteria:  Galaxies have to meet the required quality standards of the analysis pipeline (i.e. a QCFLAG field equal to zero in the {\tt pyPipe3D} VAC table; see section 4.5 of \citealt{sanchez2022} for details), and they must provide values for all of the analysed parameters. The second criterion is quite restrictive. For instance, it removes early-type systems and galaxies in general with little or no ionised gas (for which gas properties such as the oxygen abundance or the dust extinction cannot be derived). The final sample contains 3\,430 galaxies. This number is high enough to ensure the statistical significance of the results. However, we note that the sample is biased towards large star-forming galaxies.

\subsection{Random forest regressor}\label{sec:RF}

We applied a supervised machine-learning technique known as {\it random forest} (RF) regressor \citep[][]{breiman2001}. Through a combination of decision trees, the algorithm finds the input features (the galaxy properties in our case) that contain more information on the target feature ($Z_g$) and creates a model to predict it by defining a set of conditions on the values of the input features. RF regressors have been extensively used in astronomy in the past decade with high levels of success \citep[e.g.][among many others]{carliles2010, carrasco2013, sanchezmenguiano2019, moster2021}. The algorithm was implemented using the {\tt scikit-learn} package for Python \citep{pedregosa2011}. We briefly describe the basic steps we followed to run the algorithm, but we refer to the {\tt scikit-learn} User Guide documentation for more details of the complete algorithm implementation\footnote{\url{https://scikit-learn.org/stable/modules/ensemble.html\#forest}}.

We first randomly split the sample of 3\,430 galaxies into two subsets: One set was used to create the model (training sample; 2\,572 galaxies, corresponding to two-thirds of the sample), and the other set was used to evaluate its performance (test sample; 858 galaxies, comprising the remaining one-third of the sample). For the training sample, the selected set of galaxy properties was provided to the algorithm to train the model (these are the {\it predictors}). $Z_g$ was considered the target feature that was to be predicted by the RF model (this is the {\it solution}). The algorithm requires not only the predictors, but also the solutions (called {\it labels}) to train the model. When the model was trained, we applied it to the test sample, obtaining a set of predictions for $Z_g$. The comparison of these predictions with the measured values provides an estimate of the model precision. 

Before training the model, we selected the values of the parameters in the algorithm (called hyperparameters, HP) that optimise its performance. The HPs include the number of trees in the forest ($n$), the number of randomly selected features to consider in each split ($m$), the maximum depth of the trees ($max_{depth}$), the minimum number of samples required in each split ($min_{split}$), and the minimum number of samples that remain at the end of the different decision tree branches ($min_{leaf}$). We set $m=N_{\rm par} = 148$ to fully disentangle the relative interdependences among the input parameters, allowing the algorithm to select any of them at each split. If we choose $m < N_{\rm par}$, a variable containing significant information on the target will be replaced by any correlated parameter when it is not included among the randomly selected features to consider in a split. This will falsely reduce the importance of this variable in the model. By setting $m=N_{\rm par}$, any small difference in the performance of two correlated parameters will be detected by the RF, breaking possible degeneracies in the relations of the parameters. We refer to \citet[][appendix B.2]{bluck2022} for detailed tests of how well RF can perform with inter-correlated data. Nevertheless, this decision must be balanced with a careful choice of the other HPs to avoid overfitting the data. In order to select the remaining HPs, we used the fivefold cross-validation (CV) method \citep{james2013}, which consists of splitting the training sample into five subsets, called folds. We trained the model on a different combination of four folds and evaluated it on the remaining fold. The reported performance measure was then the average of the values computed in the loop. We performed 100 iterations of the entire \mbox{5-Fold} CV process, each time using different values for the HPs. We selected the HP values that achieved the highest average performance across the five folds. In our case, this resulted in $n = 700$, $max_{depth} = 30$, $min_{split} = 4$, and $min_{leaf} = 2$. We note, however, that the results we obtained are not very sensitive to the selected values for the HPs.

\begin{figure}
\centering
\resizebox{\hsize}{!}{\includegraphics{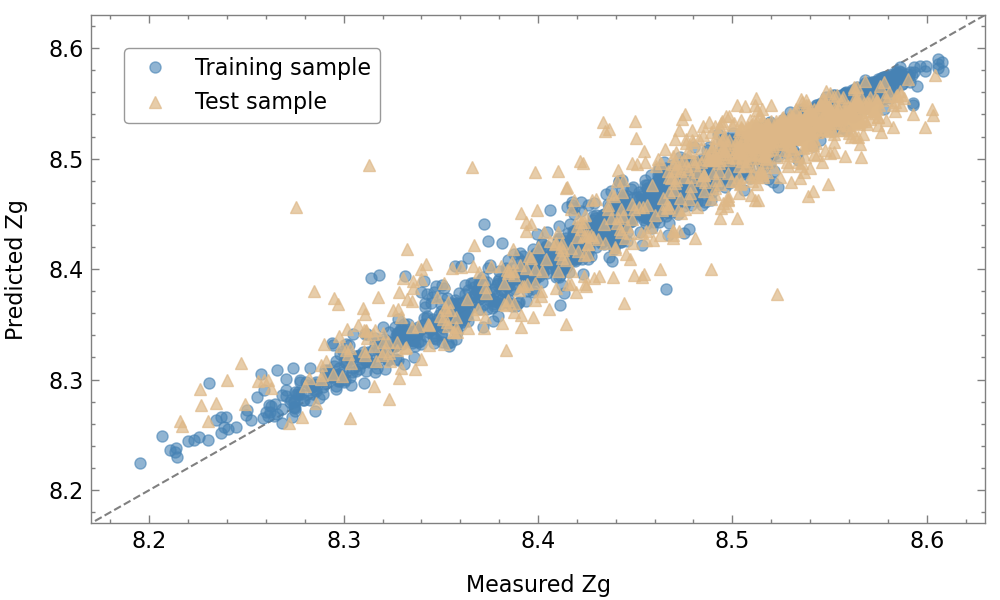}}
\caption{Global gas-phase metallicity --described by the oxygen abundance 12+log(O/H)-- predicted by the RF algorithm vs. the measured values. The blue circles correspond to the training set, and the brown triangles represent the test sample. The dashed line indicates the one-to-one relation.}
\label{fig1}
 \end{figure}

Finally, using the selected HPs, we trained the model on the full training sample, and then evaluated it on the test sample. Figure~\ref{fig1} represents the predicted $Z_g$ by the RF algorithm against the measured values for the training (blue circles) and test (brown triangles) samples. The test sample follows a trend around the one-to-one relation (dashed line), similar to the training sample ($R^2$ score of 0.87 and 0.98, respectively), showing that the algorithm is able to predict the $Z_g$ of galaxies that are not used to train the model with comparable accuracy. This ensures that the model is not affected by overfitting. The median error in the predictions (predictions-targets) is 0.017 dex, which is about the same as the observational errors propagated into $Z_g$ ($\sim$ 0.011 dex, which does not consider the systematic error in the calibrator). 

\section{Results}

\subsection{Feature importance from random forest}\label{sec:results1}

The model obtained by the RF algorithm allows us to investigate the connection between different galaxy properties (the features) and their overall gas metallicity (the target). Although the model was fed with more than 100 input parameters that characterise the galaxies, only a few of them are typically important for an accurate prediction for the target. The remaining parameters have little to no impact on $Z_g$.  

\begin{figure}
\centering
\resizebox{\hsize}{!}{\includegraphics{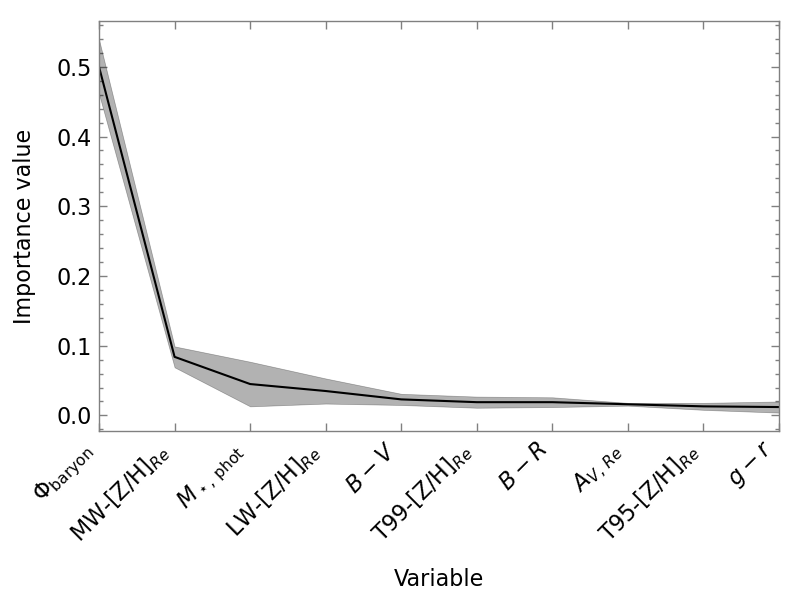}}
\caption{Average relative importance of the first ten input features in predicting the global gas metallicity for 50 runs of the RF algorithm. The features are ranked by decreasing order of importance. The shaded area represents the standard deviation of the mean relative importance among the 50 runs. The meaning of the labels is detailed in Table~\ref{tab1}.}
\label{fig2}
 \end{figure}

In order to find the most relevant features, the RF provides a measurement of the relative importance of each feature in predicting the solution. In rough outlines, the so-called feature importance is a measure of how effective the feature is at reducing variance when the variables are split along the decision trees. The higher the value, the more important the feature. Figure~\ref{fig2} shows the relative importance of the first ten input features (see Table~\ref{tab1} for a description of the labels), ranked by decreasing order of the importance value. In order to test the stability of these values, we ran the RF algorithm 50 times, each time with a different randomly selected training sample. The figure shows the average trend of the values for the 50 realisations together with the standard deviations (shaded area). To our initial surprise, the baryonic gravitational potential ($\Phi_{\rm baryon}$) was the primary factor determining the global gas metallicity, with a mean importance value of $\sim0.51$, which is well above that of the other values. The second most relevant parameter was the mass-weighted stellar metallicity measured at 1 $R_e$ (MW-[Z/H]$_{Re}$), which presents a relative importance of $\sim0.08$. The remaining properties had an importance value below 0.05 and are therefore of little or no relevance for determining the gas metallicity. This includes the stellar mass, which occupies the third position in the ranking, with a relative importance of $\sim0.04$.

The importance value of $\Phi_{\rm baryon}$ is much higher than that of the stellar mass, which means that the gravitational potential is significantly more efficient than the mass to decrease the variance when the variables are split along the decision trees. To confirm this, we ran the RF with the gravitational potential as the only parameter in the model. The root mean square error (RMSE) of the difference between the predictions (modelled $Z_g$) and the targets (measured $Z_g$) is $0.052 \pm 0.001$ (50 runs of the algorithm). If instead we use the stellar mass as the only parameter in the model, the RMSE increases up to $0.054 \pm 0.001$. This test supports the finding that $\Phi_{\rm baryon}$ better describes the gas metallicity than $M_\star$. We would also like to note that although the relative importance of the stellar mass is lower than MW-[Z/H]$_{Re}$, we must not forget the correlation between $\Phi_{\rm baryon}$ and $M_\star$, which might cause the reduced importance of $M_\star$ in the model. We also ran the RF with the stellar metallicity as the only parameter in the model. We obtained an RMSE of $0.056 \pm 0.001$, confirming that although $M_\star$ is the third most relevant parameter in the original model, it regulates the gas metallicity better than MW-[Z/H]$_{Re}$.

\begin{figure}
\centering
\resizebox{\hsize}{!}{\includegraphics{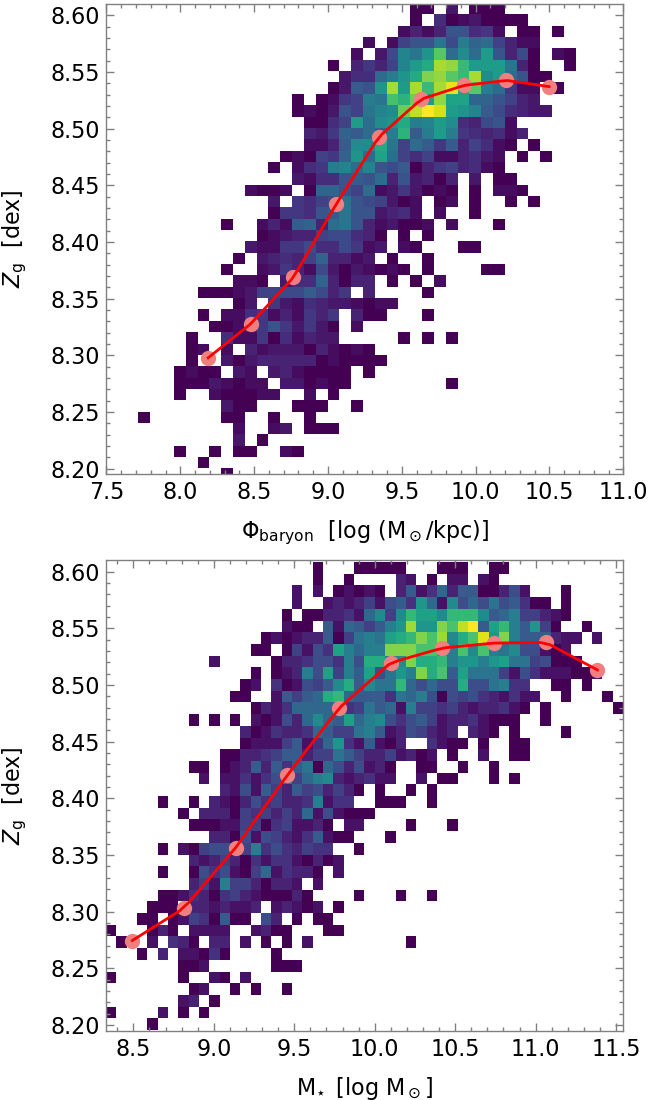}}
\caption{Global gas metallicity as a function of the stellar mass (the so-called MZR, {\it bottom}) and the galactic gravitational potential (hereafter called $\Phi$ZR, {\it top}). The brown squares represent the mean $Z_g$ values in ten mass and $\Phi$ bins, respectively. The averaged binned values were fitted with a spline function (solid red lines).} 
\label{fig4}
 \end{figure}

In order to visualise the shape of the dependence of the global gas metallicity on the stellar mass and on the baryonic gravitational potential, we present in Figure~\ref{fig4} $Z_g$ as a function of both $\Phi_{\rm baryon}$ (hereafter called $\Phi$ZR, {\it top}) and $M_{\star}$ (the so-called MZR, {\it bottom}). The shape of the relations is very similar in both cases: $Z_g$ increases when $M_{\star}$ and $\Phi_{\rm baryon}$ increase, although the flattening at the high end is more pronounced in the MZR. Regarding the dispersion, we find $\Phi$ZR to be slightly tighter than the MZR, although the difference is small ($2\%$). In any case, these results agree with the conclusions derived from the RF analysis, supporting the finding that $\Phi_{\rm baryon}$ is a slightly better predictor of $Z_g$ than $M_\star$.

\subsection{Gravitational potential, or the combined effect of stellar mass and galaxy size}\label{sec:results2}

The ratio $M_\star/R_e$ was used in the analysis as a proxy for the galactic gravitational potential at one effective radius. This first-order approximation ignores the effects of the dark matter (DM) halo on the potential in these inner regions, which is assumed to be dominated by the baryonic component (which in turn is dominated by the stellar component). We address the role of DM below.

The correlation between gas metallicity and both stellar mass and galaxy size has been described in numerous studies \citep[e.g.][]{ellison2008, sanchezalmeida2018b, sanchez2017, sanchez2019b}. Since $\Phi_{\rm baryon}$ is estimated as $M_\star/R_e$, the fact that $\Phi_{\rm baryon}$ resulted as the primary factor determining $Z_g$ could simply reflect therefore the combined effect of both stellar mass and galaxy size on $Z_g$. In this case, the combination of stellar mass and radius that best fits the metallicity could be in the form of $M_\star/R_e^{\,\alpha}$, with the exponent $\alpha$ similar to but not necessarily equal to one. In order to further test this hypothesis, we ran the RF again, and this time included 20 new input parameters in the shape of $M_\star / R_e^{\,\alpha}$, with $\alpha=0.1,..., 2$, in steps of 0.1 (\mbox{$\alpha=1$}, corresponding to $\Phi_{\rm baryon}$, and \mbox{$\alpha=2$} to $\Sigma_{\star}$). If another value of $\alpha \neq 1$ fits better than $\Phi_{\rm baryon}$, it could suggest that the real predictor of the $Z_g$ is the combined effect of stellar mass and galaxy size and not the baryon gravitational potential of the galaxy alone. 

The outcome of this test is shown in Table~\ref{tab2}, that is, the average importance values (out of 50 realisations) of the new $M_\star / R_e^{\,\alpha}$ features included in the RF, together with their position in the ranking list. Only the first five parameters are displayed. Interestingly, $\Phi_{\rm baryon}$ (\mbox{$\alpha=1$}) is replaced by $M_\star / R_e^{\,0.6}$ as the parameter with the strongest effect on $Z_g$.

\begin{table}[h]
\caption{Relative importance of the $M_\star / R_e^{\,\alpha}$ features in the RF model.}
\label{tab2}      
\centering          
\begin{tabular}{c@{\hspace{1cm}}c@{\hspace{1cm}}c}   
$\alpha$ & Importance & Position in\\
 &  value & ranking\\
\hline\\[-0.355cm]\hline\\[-0.3cm]
$0.6$ & 0.217 & 1\\[0.1cm]
$0.7$ & 0.137 & 2\\[0.1cm]
$0.8$ & 0.072 & 3\\[0.1cm]
$0.5$ & 0.071 & 4\\[0.1cm]
$0.4$ & 0.020 & 7\\[0.1cm]
\hline                  
\end{tabular}
\tablefoot{$\alpha$ ranges from 0.1 to 2 in steps of 0.1. These values are averaged over 50 runs of the algorithm. The features are ranked by decreasing order of the importance value. Only the first five parameters are shown. See Sec.~\ref{sec:results2} for more details.}
\end{table}

This result seems to suggest that the gravitational potential set by the baryons is indeed not the best predictor of $Z_g$, but a particular combination of mass and size, $M_\star / R_e^{\,0.6}$. However, as we pointed out above, in our approximation, we ignored the effects of the DM halo. $\alpha\simeq 0.6$ might account for the inclusion of the DM component in the whole gravitational potential of the galaxy. In this way, $M_\star / R_e^{\,0.6}$ would be a better tracer of the real $\Phi$ than $M_\star / R_e$, and our main conclusion would remain: The total gravitational potential would be the galactic property with the strongest impact on $Z_g$. This seems indeed to be the case, as we discuss below.

  In order to investigate the conjecture, we worked out the expression for the fraction of the DM,
  \begin{equation}
 f_{DM} = M_{\rm DM} / (M_{\rm baryon}+ M_{\rm DM}),
\end{equation}
that is needed for the logarithm of the total potential to scale with the parameter used in the RF, that is, the logarithm of $M_\star /R_e^{\,\alpha}$, 
\begin{equation}
  \log\Phi = \log(M_T / R_e) = a+b\,\log(M_\star /R_e^{\,\alpha}),
  \label{eq:phi0}
\end{equation}
where $a$ and $b$ are two scaling constants, and $M_T$ is the total mass, composed of baryons, $M_{\rm baryon}$, and DM, $M_{\rm DM}$. Assuming a functional dependence of $f_{\rm DM}$ of the type
\begin{equation}
  f_{\rm DM} = 1- B \, (M_\star/2\pi R_e^{\,2})^\beta,
  \label{eq:phi1}
\end{equation}
with $B$ an arbitrary factor, we find after some algebra that the functional form in Eq.~(\ref{eq:phi1}) is consistent with Eq.~(\ref{eq:phi0}) provided 
\begin{equation}
  \beta=(1-\alpha)/(2-\alpha).
  \label{eq:phi2}
\end{equation}
We considered $M_{\rm baryon}\simeq M_\star$, which is a good approximation for most MaNGA galaxies \citep{lin2020}. The functional form in Eq.~(\ref{eq:phi1}) was chosen because it seems to be followed by the galaxies produced in cosmological numerical simulations, as we argue below.

\begin{figure}
\centering
\resizebox{\hsize}{!}{\includegraphics{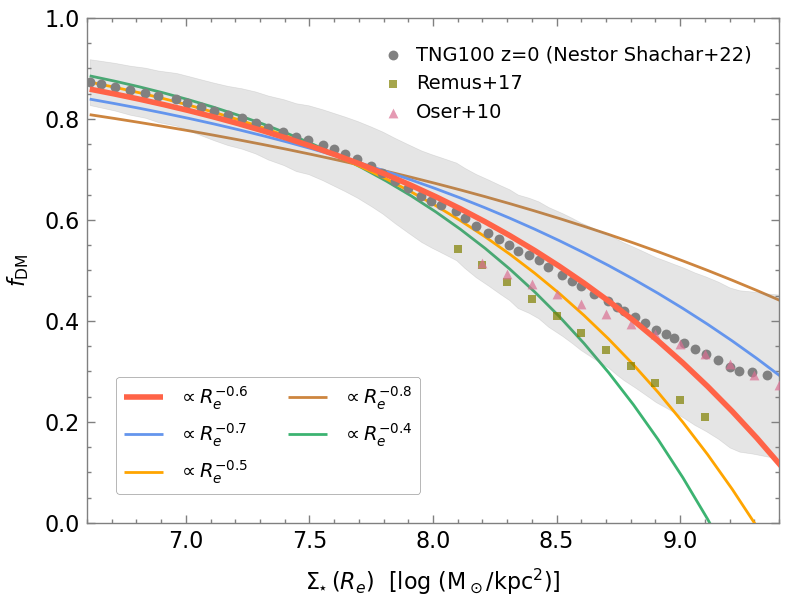}}
\caption{Dependence of the dark matter fraction on the baryonic surface density for galaxies in the local Universe predicted by Illustris-TNG100 simulations \citep[][figure~5]{nestorshachar2023}. This relation is compared with the one obtained from the five $M_\star / R_e^{\,\alpha}$ parameters with the highest relative importance in the RF. $M_\star / R_e^{\,0.6}$ provides a good fit for the theoretical relation. Two additional relations for the dependence of $f_{\rm DM}$ on $\Sigma_{\star}$ from different simulations are also included \citep{oser2010, remus2017}.} 
\label{fig5}
 \end{figure}

Figure~\ref{fig5} uses Eqs.~(\ref{eq:phi1}) and (\ref{eq:phi2}) to show the baryon fraction for Eq.~(\ref{eq:phi0}) to hold, depending on the value of $\alpha$ (the coloured lines). The figure also includes the theoretical expectation for $f_{\rm DM}$ (the symbols) as predicted by the Illustris-TNG100 simulations \citep{lovell2018}  at redshift 0, a prediction that we extracted from fig.~5 in \citet[][]{nestorshachar2023}. The predictions from two additional simulations by \citet{oser2010} and \citet{remus2017} are also shown, reflecting that the slope of the relation between $f_{\rm DM}$ and $\Sigma_{\star}$ given by $\beta$ might slightly change depending on the set of simulations that is used.
Figure~\ref{fig5} illustrates that $\alpha=0.6$ (the solid orange line) matches the theoretical expectation from Illustris-TNG100 (grey symbols and shaded area) extremely well, except for extreme surface densities. The match is encouraging because no freedom or fitting is shown in the representation in the figure (except for a shift in the y-axis given by the constant $B$). In any case, the match for $\alpha=0.6$ is much better than 0.5 or 0.7. We interpret this match as evidence that $M_\star / R_e^{\,0.6}$ traces $\Phi$, and therefore, evidence that our RF reveals a true relation between the total gravitational potential of the galaxy and the gas metallicity, which is tighter and more primordial than the MZR or the baryonic $\Phi$ZR.

\section{Discussion}\label{sec:dis}

The search for relations between integrated physical properties in galaxies has been widely addressed in the literature. In particular, scaling relations involving the gas-phase oxygen abundance (as a proxy for the gas metalllicity, $Z_g$) have received much attention because it is a tracer of the overall chemical evolution of galaxies, modulated by the inflow or outflow of gas in different regions \citep{sanchezalmeida2014, maiolino2019, sanchez2020}. For this reason, the dependence of $Z_g$ on other properties (or vice versa) can provide important insights into galaxy formation and evolution. 

In this study, we analysed the global properties of a sample of more than 3000 nearby galaxies from the SDSS-IV MaNGA survey \citep{bundy2015}. The aim of this analysis is to evaluate the galactic property that is the best predictor of $Z_g$, and to do this, we explored the effect of 148 parameters extracted from the {\tt pyPipe3D} Value Added catalogue \citep[][]{sanchez2022}. We include the parameters that were previously reported to strongly correlate with $Z_g$, such as $M_{\star}$, $\Phi$, $\Sigma_{\star}$, and $M_{\rm gas}$. We used the oxygen abundance measured at 1 $R_e$ as a proxy for the global gas metallicity, determined with the O3N2 calibration proposed by \citet{marino2013}. 

For this analysis, we applied a RF regressor, which is a supervised machine-learning technique that has been proven successful in the past in searches for the primary relations between galaxy properties \citep[e.g.][]{sanchezmenguiano2019, baker2023b, baker2023a}. Our trained model was able to predict $Z_g$ with an average error (defined as the difference between the predictions and the targets) of about the propagated uncertainties associated with the measured values.

Based on the relative importance of each galaxy property in predicting $Z_g$, we obtained that the baryonic gravitational potential (defined as $\Phi_{\rm baryon} = M_{\star} / R_e$) is the most relevant parameter of the model, followed by MW-[Z/H]$_{Re}$ and $M_{\star}$, with an importance value more than five times lower than that of $\Phi_{\rm baryon}$. This result was supported by the outcome of a subsequent run of the RF with only one parameter in the model. When that parameter was $M_{\star}$ (MW-[Z/H]$_{Re}$), the RMSE of the differences between the predictions and the targets was $0.054 \, (0.056) \pm 0.001$, whereas this value was reduced to $0.052 \pm 0.001$ for $\Phi_{\rm baryon}$. The difference between these values and that found for the complete model (0.029) suggests that, assuming there is no overfitting, other secondary parameters play a significant role in shaping the gas metallicity. This will be investigated in a future study (S\'anchez-Menguiano et al. in prep.). Furthermore, the scatter of the relations between $Z_g$ and $\Phi_{\rm baryon}$ ($\Phi$ZR) and between $Z_g$ and $M_{\star}$ (the so-called MZR) shows that the $\Phi$ZR is slightly tighter than the MZR ($2\%$). Although the differences are small in both cases, these tests support the finding that $\Phi_{\rm baryon}$ is better suited than $M_\star$ to describe the gas metallicity.

These results are representative of the whole population of analysed galaxies, which spreads over a wide range of stellar masses (8.5 < $\log(M_{\star}/M_{\odot})$ < 11.5). Although our mass coverage is not homogeneous (there is an excess of massive galaxies compared to low-mass systems), we show in Appendix~\ref{sec:appendix3} that this bias does not affect the analysis because using a subsample of galaxies with a uniform mass distribution (constant number of objects per mass bin) leads to the same results. Restricting the analysis to a more limited mass range, however, might lead to different conclusions. For instance, because both the MZR and the $\Phi$ZR saturate at high masses, a sample that is highly dominated by massive systems and with a poor coverage of low-mass galaxies could reduce the effect of both $M_\star$ and $\Phi_{\rm baryon}$ on $Z_g$ (because $Z_g$ remains almost constant, while both parameters increase). It is therefore important to rely on a wide mass range that allows covering a wide dynamical range for $Z_g$ in order to detect true primary and secondary relations.

It is interesting to note that the stellar mass estimated from photometry (within the MaNGA FoV) based on the use of the $M/L$ ratio and the relations provided in \citet{bell2000} is a better predictor of $Z_g$ than the mass derived from the spectroscopic analysis of the stellar populations performed by {\tt pyPipe3D}. Whereas the first obtains an importance value of $0.043$, the later obtains just $0.001$. Another estimation of the integrated stellar mass from the NSA catalog provides an intermediate value ($0.010$). Photometric stellar masses are considered to be less accurate, but are probably more precise than estimates based on the spectroscopic analysis performed by {\tt pyPipe3D}. For a more detailed comparison of the mass estimations, we refer to \citet{sanchez2022}.

Due to the well-known discrepancies between the oxygen abundances derived using different strong-line calibrators \citep[for a review on the topic, see][]{kewley2019}, we reproduced the analysis employing 18 alternative O/H estimators. Consistent with the use of O3N2-M13, in 15 out of the 18 cases, we obtained that $\Phi_{\rm baryon}$ is the best predictor of the global gas metallicity. This strengthens the validity of the result, which is independent of the method used to estimate the abundances. In the remaining three, the discrepancies might be due to the small dynamical range of the calibrators (more details can be found in Appendix~\ref{sec:appendix2}).

This is not the first study to find that the correlation of the gas metallicity with the baryonic gravitational potential is stronger than with the stellar mass. Using a volume-limited sample drawn from the publicly available SDSS DR7, \citet{deugenio2018} analysed the relation between $Z_g$ and three galaxy properties: $\Phi_{\rm baryon}$, $M_{\star}$, and average $\Sigma_{\star}$. To minimise aperture bias, the authors also defined three aperture-matched subsamples to ensure an homogeneous coverage of the physical size of galaxies from the fixed-aperture SDSS fibres: $R_{\rm fib}/R_e =$ 0.5, 1.0, and 1.5. For the three subsamples, they find that $\Phi$ZR had the lowest scatter and the highest Spearman correlation coefficient. In addition, this relation had the lowest residual trends with galaxy size. Based on these results, the authors concluded that of the three parameters, $\Phi_{\rm baryon}$ is the best predictor of $Z_g$. According to \citet{deugenio2018}, there are two possible explanations for this. The correlation of $Z_g$ and $\Phi$ might arise from the combination of metal loss (which depends on the value of the escape velocity, which in turn is related to $M_{\star}$) and metal enrichment (driven by $f_{\rm gas}$, and therefore $\Sigma_{\star}$). Each of these effects induces a correlation between $Z_g$ and $M_{\star}/R_e^{\,n}$, with $n=0$ and $n=2$, respectively, and their combination could produce an overall correlation with an intermediate value of $n$. However, because the metallicity of low-mass galaxies is highly affected by outflows, $Z_g$ is predicted to correlate best with $M_{\star}$ at the low-mass end of the relation. Similarly, because the escape fraction for high-mass galaxies is small, $Z_g$ should correlate best with $\Sigma_{\star}$ at the high-mass end of the relation. These predicted trends are not observed in their analysis, and therefore, the authors pointed to another explanation for the $\Phi$MR: a direct physical link between the average depth of the gravitational potential and the average metallicity, which is a consequence of an existing local relation between $Z_g$ and the escape velocity (this local quantity is directly related with the global gravitational potential). We note that this local relation does exist and has already been reported in \citealt{barreraballesteros2018}. The first explanation would imply that the $\Phi$MR is the result of the already known MZR and $\Sigma$ZR, whereas the latter indicates that the potential–metallicity relation is the primordial relation, and the MZR and the $\Sigma$ZR would arise from the local relation between $Z_g$ and $\Phi$.

The first explanation provided by \citet{deugenio2018} could be further tested by re-applying the RF and include different parameters in the model in the shape of $M_\star / R_e^{\,\alpha}$. We tried 20 combinations with $\alpha=0.1,..., 2$, in steps of 0.1. Interestingly, $\Phi_{\rm baryon}$ (\mbox{$\alpha=1$}) is replaced by $M_\star / R_e^{\,0.6}$ as the parameter with the strongest effect on $Z_g$. When the alternative calibrations were applied to determine $Z_g$, the coefficients $\alpha=0.7$ and $\alpha=0.8$ in some cases replaced $\alpha=0.6$ as those occupying the highest position in the importance ranking. This may suggest that the $\Phi$MR is indeed not the primordial relation, but results from the MZR and $\Sigma$ZR with a particular combination of the effects of metal loss and metal enrichment.

However, we provide here another explanation to justify the result. We have ignored the effects of the DM halo in the analysis so far. $\alpha\simeq 0.6$ might account for the inclusion of the DM component in the whole gravitational potential of the galaxy. In Figure~\ref{fig5} we showed that a scale of the form $M_\star / R_e^{\, 0.6}$ for the total gravitational potential matches the theoretical relation between the DM fraction and the baryonic surface density predicted in the Illustris TNG cosmological numerical simulation very well \citep{nestorshachar2023}. In this way, $M_\star / R_e^{\,0.6}$ would be a better tracer of the real $\Phi$ than $M_\star / R_e$, and therefore, the total gravitational potential (and not its baryonic component) would be the galactic property that predicts $Z_g$ best. The substitution of $\alpha=0.6$ by $\alpha=0.7$ or $\alpha=0.8$ for the parameter with the strongest effect on $Z_g$ when using other abundance calibrations can be accounted for by the large scatter in the prediction by \citet{nestorshachar2023} (shaded area in Fig.~\ref{fig5}), and the fact that the slope of the relation might vary slightly depending on the simulations used \citep{oser2010, remus2017}. 

A recent study by \citet{baker2023c} also explored the relation between $Z_g$ and (i) $M_{\star}$, (ii) dynamical mass ($M_{\rm dyn}$, DMZR), and (iii) gravitational potential (estimated as $\Phi = M_{\rm dyn}/R_{e}$) using MaNGA data. Based on three different methods (average dispersion, partial correlation coefficients, and RF), the authors reported that the gas metallicity primarily depends on the stellar mass, with little or no dependence on either dynamical mass or gravitational potential when the dependence on stellar mass is taken into account. However, their $M_{\rm dyn}$ were derived via Jeans anisotropic modelling, which is a complex method with its own systematic problems that might increase the intrinsic scatter in DMZR and $\Phi$ZR with respect to MZR \citep[e.g.][]{li2016, elbadry2017, read2021}. We argue that our derivation of $\Phi_{\rm baryon}$, although less accurate, produces more precise results that allow us to determine the intrinsic scatter in $\Phi$ZR, resulting in a tighter relation than the MZR. This is in contrast to the results reported by \citet{baker2023c}.

\section{Conclusions}\label{sec:concl}

We studied the dependence of the gas-phase metallicity ($Z_g$) on a large set of galaxy properties, including parameters that have been reported in the literature to correlate strongly with $Z_g$, such as the stellar mass ($M_{\star}$), the baryonic gravitational potential ($\Phi_{\rm baryon}$), the average stellar mass surface density ($\Sigma_{\star}$), or the gas mass ($M_{\rm gas}$). In order to do this, we applied a RF regressor algorithm on a sample of more than 3000 nearby galaxies from the SDSS-IV MaNGA survey. Our results are listed below.\\[-0.2cm]

      (i) $M_{\rm \star}$/$R_e$, as a proxy for $\Phi_{\rm baryon}$, is found to be the primary factor determining $Z_g$ (Section~\ref{sec:results1}; Figure~\ref{fig2}). Because the gravitational potential regulates the ability of a galaxy to retain metals, it is not surprising that $\Phi$MR is the primordial scaling relation involving $Z_g$.
       
      (ii) When we included a parameter of the type $M_\star / R_e^{\,\alpha}$ in the analysis, we find that $\alpha=0.6$ best correlates with $Z_g$ (Section~\ref{sec:results2}; Table~\ref{tab2}). We explored the effect of the DM halo on the gravitational potential, and we showed that a scale of the form $M_\star / R_e^{\, 0.6}$ for the total $\Phi$ matches the theoretical relation between the DM fraction and the baryonic surface density predicted in cosmological numerical simulations of galaxy formation very well (Section~\ref{sec:results2}; Figure~\ref{fig5}). This suggests that the total gravitational potential would be the galactic property that predicts $Z_g$ best.
      
      (iii) In the analysis, we used the oxygen abundance measured with the O3N2 calibration by \citet{marino2013} as an indicator of the gas metallicity. The use of alternative estimators still gives $\Phi$MR as the primordial relation (Appendix~\ref{sec:appendix2}). The inclusion of the DM halo might affect the particular coefficient $\alpha$, which could be better described by theoretical predictions from other simulations (Section~\ref{sec:dis}; Figure~\ref{fig5}).
      
     (iv) We find evidence that $\Phi_{\rm baryon}$ alone cannot predict $Z_g$ and that other secondary parameters might play a significant role in shaping the gas metallicity (Section~\ref{sec:dis}). This will be investigated in a future study (S\'anchez-Menguiano et al. in prep.).

\begin{acknowledgements}
We thank the anonymous referee for suggestions that allowed us to improve the paper. LSM acknowledges support from Juan de la Cierva fellowship (IJC2019-041527-I) and from project PID2020-114414GB-100, financed by MCIN/AEI/10.13039/501100011033. This research was partly funded by the Spanish Ministry of Science and Innovation, projects PID2019-107408GB-C43 and PID2022-136598NB-C31(ESTALLIDOS), and by Gobierno de Canarias through EU FEDER funding, project PID2020010050. S.F.S. thanks the PAPIIT-DGAPA AG100622 project.\\[0.1cm]
This project makes use of the MaNGA-Pipe3D dataproducts. We thank the IA-UNAM MaNGA team for creating this catalogue, and the Conacyt Project CB-285080 for supporting them. 
\end{acknowledgements}

\bibliographystyle{aa} 
\bibliography{bibliography} 

\begin{appendix} 

\section{Galaxy parameters}\label{sec:appendix1}
In this appendix, we list all the galactic parameters extracted from the {\tt pyPipe3D} VAC that are included in the RF analysis. This includes a brief description of their meaning. The information is displayed in Table~\ref{tabA1}. 

\begin{table*}
\begin{center}
\caption{Parameters extracted from the {\tt pyPipe3D} VAC included in the analysis.}
\begin{tabular}{lll}
\hline\hline\\[-0.3cm]
Name & Unit & Description \\
\hline\\[-0.3cm]
mangaid & -- & MaNGA ID \\ 
log\_SFR\_H$\alpha$& log(M$_\odot$ yr$^{-1}$) & Integrated star-formation rate derived from the integrate H$\alpha$ luminosity \\ 
FoV & -- & Ratio between the diagonal radius of the cube and Re \\
Re\_kpc & kpc & Effective radius in kpc \\ 
log\_Mass & log(M$_\odot$) & Integrated stellar mass in units of the solar mass in logarithm scale \\ 
log\_SFR\_ssp & log(M$_\odot$ yr$^{-1}$) & Integrated SFR derived from the SSP analysis t$<$32Myr \\ 
ZH\_LW\_Re\_fit & dex & LW metallicity of the stellar population at Re, normalised to the solar value, \\
& & in logarithm scales \\ 
alpha\_ZH\_LW\_Re\_fit & dex/Re & Slope of the gradient of the LW metallicity of the stellar population \\ 
ZH\_MW\_Re\_fit & dex & MW metallicity of the stellar population at Re, normalised to the solar value, \\
& & in logarithm scales \\ 
alpha\_ZH\_MW\_Re\_fit & dex/Re & Slope of the gradient of the MW log-metallicity of the stellar population \\ 
Age\_LW\_Re\_fit & log(yr) & Luminosity weighted age of the stellar population in logarithm scale \\ 
alpha\_Age\_LW\_Re\_fit & log(yr)/Re & Slope of the gradient of the LW log-age of the stellar population \\ 
Age\_MW\_Re\_fit & log(yr) & Mass weighted age of the stellar population in logarithm \\ 
alpha\_Age\_MW\_Re\_fit & {log(yr)}/Re & Slope of the gradient of the MW log-age of the stellar population \\ 
Re\_arc & arcsec & Adopted effective radius in arcsec \\ 
DL & { Mpc} & Adopted luminosity distance \\ 
DA & { Mpc} & Adopted angular-diameter distance \\ 
PA & degrees & Adopted position angle in degrees \\ 
ellip & -- & Adopted ellipticity \\ 
log\_Mass\_gas & log(M$_\odot$) & Integrated gas mass in units of the solar mass in logarithm scale \\ 
vel\_sigma\_Re & -- & Velocity/dispersion ratio for the stellar populations within 1.5 Re \\ 
log\_SFR\_SF & log(M$_\odot$ yr$^{-1}$) & Integrated SFR using only the spaxels compatible with SF \\ 
log\_SFR\_D\_C & log(M$_\odot$ yr$^{-1}$) & Integrated SFR diffuse corrected \\ 
Sigma\_Mass\_cen & log(M$_\odot$ pc$^{-2}$) & Stellar Mass surface density in the central aperture \\ 
Sigma\_Mass\_Re & log(M$_\odot$ pc$^{-2}$) & Stellar Mass surface density at 1Re \\ 
Sigma\_Mass\_ALL & log(M$_\odot$ pc$^{-2}$) & Average Stellar Mass surface density within the entire FoV \\ 
T30 & Gyr & Lookback time at which the galaxy has formed 30\% of its stellar mass \\ 
ZH\_T30 & dex & Stellar metallicity, normalised by the solar value, in logarithm-scale at T30   time \\ 
ZH\_Re\_T30 & dex & Stellar metallicity, normalised by the solar value, in logarithm-scale at Re at T30   time \\ 
a\_ZH\_T30 & dex/Re & Slope of the ZH gradient at T30   time \\ 
T40 & Gyr & Lookback time at which the galaxy has formed 40\% of its stellar mass \\ 
ZH\_T40 & dex & Stellar metallicity, normalised by the solar value, in logarithm-scale at T40   time \\ 
ZH\_Re\_T40 & dex & Stellar metallicity, normalised by the solar value, in logarithm-scale at Re at T40   time \\ 
a\_ZH\_T40 & dex/Re & Slope of the ZH gradient at T40   time \\ 
T50 & Gyr & Lookback time at which the galaxy has formed 50\% of its stellar mass \\ 
ZH\_T50 & dex & Stellar metallicity, normalised by the solar value, in logarithm-scale at T50   time \\ 
ZH\_Re\_T50 & dex & Stellar metallicity, normalised by the solar value, in logarithm-scale at Re at T50   time \\ 
a\_ZH\_T50 & dex/Re & Slope of the ZH gradient at T50   time \\ 
T60 & Gyr & Lookback time at which the galaxy has formed 60\% of its stellar mass \\ 
ZH\_T60 & dex & Stellar metallicity, normalised by the solar value, in logarithm-scale at T60   time \\ 
ZH\_Re\_T60 & dex & Stellar metallicity, normalised by the solar value, in logarithm-scale at Re at T60   time \\ 
a\_ZH\_T60 & dex/Re & Slope of the ZH gradient at T60   time \\ 
T70 & Gyr & Lookback time at which the galaxy has formed 70\% of its stellar mass \\ 
ZH\_T70 & dex & Stellar metallicity, normalized by the solar value, in logarithm-scale at T70   time \\ 
ZH\_Re\_T70 & dex & Stellar metallicity, normalized by the solar value, in logarithm-scale at Re at T70   time \\ 
a\_ZH\_T70 & dex/Re & Slope of the ZH gradient at T70   time \\ 
T80 & Gyr & Lookback time at which the galaxy has formed 80\% of its stellar mass \\ 
ZH\_T80 & dex & Stellar metallicity, normalized by the solar value, in logarithm-scale at T80   time \\ 
ZH\_Re\_T80 & dex & Stellar metallicity, normalized by the solar value, in logarithm-scale at Re at T80   time \\ 
a\_ZH\_T80 & dex/Re & Slope of the ZH gradient at T80   time \\ 
T90 & Gyr & Lookback time at which the galaxy has formed 90\% of its stellar mass \\ 
ZH\_T90 & dex & Stellar metallicity, normalized by the solar value, in logarithm-scale at T90   time \\ 
ZH\_Re\_T90 & dex & Stellar metallicity, normalized by the solar value, in logarithm-scale at Re at T90   time \\ 
a\_ZH\_T90 & dex/Re & Slope of the ZH gradient at T90   time \\ 
T95 & Gyr & Lookback time at which the galaxy has formed 95\% of its stellar mass \\ 
\end{tabular}\label{tabA1}
\tablefoot{The MaNGA ID is included in the table only for identification. We refer to \citet{sanchez2022} for more details of the derivation of the parameters.}
\end{center}
\end{table*}

\addtocounter{table}{-1}

\begin{table*}
\begin{center}
\caption{continued.}
\begin{tabular}{lll}
\hline\hline\\[-0.3cm]
Name & Unit & Description \\
\hline\\[-0.3cm]
ZH\_T95 & dex & Stellar metallicity, normalized by the solar value, in logarithm-scale at T95   time \\ 
ZH\_Re\_T95 & dex & Stellar metallicity, normalised by the solar value, in logarithm-scale at Re \\
& & at T95   time \\ 
a\_ZH\_T95 & dex/Re & Slope of the ZH gradient at T95   time \\ 
T99 & Gyr & Lookback time at which the galaxy has formed 99\% of its stellar mass \\ 
ZH\_T99 & dex & Stellar metallicity, normalised by the solar value, in logarithm-scale at T99 time \\ 
ZH\_Re\_T99 & dex & Stellar metallicity, normalised by the solar value, in logarithm-scale at Re \\
& & at T99   time \\ 
a\_ZH\_T99 & dex/Re & Slope of the ZH gradient at T99   time \\ 
log\_Mass\_gas\_Av\_gas\_OH & log(M$_\odot$) & Integrated gas mass using the calibrator log\_Mass\_gas\_Av\_gas\_OH, \\
& & in logarithm-scale\\ 
log\_Mass\_gas\_Av\_ssp\_OH & log(M$_\odot$) & Integrated gas mass using the calibrator log\_Mass\_gas\_Av\_ssp\_OH, \\
& & in logarithm-scale\\ 
vel\_ssp\_2 & km s$^{-1}$ & Stellar velocity at 2 Re \\ 
vel\_Ha\_2 & km s$^{-1}$ & H$\alpha$ velocity at 2 Re \\ 
vel\_ssp\_1 & km s$^{-1}$ & Stellar velocity at 1 Re \\ 
vel\_Ha\_1 & km/s & H$\alpha$ velocity at 1 Re \\ 
log\_SFR\_ssp\_100Myr & log(M$_\odot$ yr$^{-1}$) & Integrated SFR derived from the SSP analysis for t$<$100Myr \\ 
log\_SFR\_ssp\_10Myr & log(M$_\odot$ yr$^{-1}$) & Integrated SFR derived from the SSP analysis for t$<$10Myr \\ 
vel\_disp\_Ha\_cen & km/s & H$\alpha$ Velocity dispersion at the central aperture \\ 
vel\_disp\_ssp\_cen & km/s & Stellar velocity dispersion at the central aperture \\ 
vel\_disp\_Ha\_1Re & km/s & H$\alpha$ velocity dispersion at 1 Re \\ 
vel\_disp\_ssp\_1Re & km/s & Stellar velocity dispersion at 1 Re \\ 
log\_Mass\_in\_Re & log(M$_\odot$) & Integrated stellar mass within one optical Re \\ 
ML\_int & M$_\odot$/L$_\odot$ & V-band mass-to-light ratio from integrated quantities \\ 
ML\_avg & M$_\odot$/L$_\odot$ & V-band mass-to-light ratio averaged across the FoV \\ 
R50\_kpc\_V & kpc & {Radius at which it is integrated half of the light in the V-band within the FoV} \\ 
R50\_kpc\_Mass & kpc & {Radius at which is integrated half of the stellar mass within the FoV } \\ 
log\_Mass\_corr\_in\_R50\_V & log(M$_\odot$) & Integrated stellar mass within { R50\_kpc\_V} \\ 
log\_Mass\_gas\_Av\_gas\_log\_log & log(M$_\odot$) & Molecular gas mass derived from the dust extinction { using the log-log calibration} \\ 
Av\_gas\_Re & mag & Dust extinction in the V-band derived from the H$\alpha$/H$\beta$ line ratio { at 1 Re} \\ 
Av\_ssp\_Re & mag & Dust extinction in the V-band derived from the analysis of the stellar population \\ 
Lambda\_Re & -- & Specific angular momentum ($\lambda$-parameter) for the stellar populations within 1 Re\\ 
nsa\_redshift & -- & Redshift derived by the NSA collaboration \\ 
nsa\_mstar &  log(M$_\odot$) & Stellar Mass derived by the NSA collaboration \\ 
nsa\_inclination & deg & Inclination derived by the NSA collaboration \\ 
O/H\_Re & dex & Oxygen abundance at the effective radius ($R_e$) using the M13 O3N2-calibrator or \\
& & one of the alternative calibrators described in Table~\ref{tabB1}\\
OH\_alpha\_fit & dex/Re & Slope of the O/H gradient using the same calibrator adopted for the \\
& & oxygen abundance determination \\ 
Ne\_Oster\_S\_Re\_fit & dex & Electron density using the Oster\_S estimator at 1Re \\ 
Ne\_Oster\_S\_alpha\_fit & dex/Re & Slope of the n\_e gradient using the Oster\_S estimator \\ 
Hd\_Re\_fit & \AA\ & Value of the Hd stellar index at 1Re \\
Hd\_alpha\_fit & \AA/Re & Slope of the gradient of the Hd index \\ 
Hb\_Re\_fit & \AA\ & Value of the H$\beta$ stellar index at 1Re \\ 
Hb\_alpha\_fit & \AA/Re & Slope of the gradient of the H$\beta$ index \\ 
Mgb\_Re\_fit & \AA\ & Value of the Mgb stellar index at 1Re \\ 
Mgb\_alpha\_fit & \AA/Re & Slope of the gradient of the Mgb index \\ 
Fe5270\_Re\_fit & \AA\ & Value of the Fe5270 stellar index at 1Re \\ 
Fe5270\_alpha\_fit & \AA/Re & Slope of the gradient of the Fe5270 index \\ 
Fe5335\_Re\_fit & \AA\ & Value of the Fe5335 stellar index at 1Re \\ 
Fe5335\_alpha\_fit & \AA/Re & Slope of the gradient of the Fe5335 index \\ 
D4000\_Re\_fit1 & -- & Value of the D4000 stellar index at 1Re \\ 
D4000\_alpha\_fit & -- & Slope of the gradient of the D4000 index \\ 
Hdmod\_Re\_fit & \AA\ & Value of the Hdmod stellar index at 1Re \\
Hdmod\_alpha\_fit & \AA/Re & Slope of the gradient of the Hdmod index \\ 
Hg\_Re\_fit & \AA\ & Value of the Hg stellar index at 1Re \\ 
Hg\_alpha\_fit & \AA/Re & Slope of the gradient of the Hg index \\ 
\end{tabular}
\end{center}
\end{table*}

\addtocounter{table}{-1}

\begin{table*}
\begin{center}
\caption{continued.}
\begin{tabular}{lll}
\hline\hline\\[-0.3cm]
Name & Unit & Description \\
\hline\\[-0.3cm]
u\_band\_mag & mag & u-band magnitude derived from the original cube \\ 
u\_band\_abs\_mag & mag & u-band absolute magnitude derived from the original cube \\ 
g\_band\_mag & mag & g-band { magnitude derived from the original cube} \\ 
g\_band\_abs\_mag & mag & g-band abs. { magnitude derived from the original cube} \\ 
r\_band\_mag & mag & r-band { magnitude derived from the original cube} \\ 
r\_band\_abs\_mag & mag & r-band abs. { magnitude derived from the original cube} \\ 
i\_band\_mag & mag & i-band { magnitude derived from the original cube} \\ 
i\_band\_abs\_mag & mag & i-band abs. { magnitude derived from the original cube} \\ 
B\_band\_mag & mag & B-band { magnitude derived from the original cube} \\ 
B\_band\_abs\_mag & mag & B-band abs. { magnitude derived from the original cube} \\ 
V\_band\_mag & mag & V-band { magnitude derived from the original cube} \\ 
V\_band\_abs\_mag & mag & V-band abs. { magnitude derived from the original cube} \\ 
RJ\_band\_mag & mag & R-band { magnitude derived from the original cube} \\ 
RJ\_band\_abs\_mag & mag & R-band abs. { magnitude derived from the original cube} \\ 
R50 & arcsec & {Radius at which it is integrated half of the light within the FoV in the g-band} \\ 
R90 & arcsec & {Radius at which it is integrated 90\%\ of the light within the FoV in the g-band} \\ 
C & -- & R90/R50 concentration index \\
B-V & mag & B-V color \\ 
B-R & mag & B-R color \\ 
log\_Mass\_phot & log(M$_\odot$) & stellar masses derived from { the photometric data within the FoV} \\ 
V-band\_SB\_at\_Re & mag/arcsec$^2$ & V-band surface brightness at 1Re \\ 
V-band\_SB\_at\_R\_50 & mag/arcsec$^2$ & V-band surface brightness at R50 \\
nsa\_sersic\_n\_morph & -- & NSA sersic index \\ 
u-g & mag & u-g NSA color \\ 
g-r & mag & g-r NSA color \\
r-i & mag & r-i NSA color \\ 
i-z & mag & i-z NSA color \\ 
P(CD) & -- & Probability of being a CD galaxy \\ 
P(E) & -- & Probability of being a E galaxy \\ 
P(S0) & -- & Probability of being a S0 galaxy \\ 
P(Sa) & -- & Probability of being a Sa galaxy \\ 
P(Sab) & -- & Probability of being a Sab galaxy \\ 
P(Sb) & -- & Probability of being a Sb galaxy \\ 
P(Sbc) & -- & Probability of being a Sbc galaxy \\ 
P(Sc) & -- & Probability of being a Sc galaxy \\ 
P(Scd) & -- & Probability of being a Scd galaxy \\ 
P(Sd) & -- & Probability of being a Sd galaxy \\ 
P(Sdm) & -- & Probability of being a Sdm galaxy \\ 
P(Sm) & -- & Probability of being a Sm galaxy \\ 
P(Irr) & -- & Probability of being a Irr galaxy \\ 
best\_type\_n & -- & Best morphological type index based on the NN analysis \\ 
$\Phi_{\rm baryon}$& log(M$_\odot$ kpc$^{-1}$) & Baryonic gravitational potential at Re estimated as $M_{\rm \star, \, phot}$/$R_e$\\
$\Sigma_{\star}$ & log(M$_\odot$ kpc$^{-2}$) & Average stellar mass surface density within Re estimated as $M_{\rm \star, \, phot}$/$R_e^2$\\
\hline
\end{tabular}
\end{center}
\end{table*}

\FloatBarrier

\section{Effect of the adopted metallicity calibration}\label{sec:appendix2}

The derivation of the gas metallicity involves a large number of systematics and sources of uncertainties, with well-known discrepancies between the use of different strong-line calibrators \citep[e.g.][]{kewley2008, lopezsanchez2012, kewley2019}. In this regard, we replicated the analysis employing 18 alternative calibrators for which the derivation of the oxygen abundance is provided in the {\tt pyPipe3D} VAC table to assess whether the choice of the gas-metallicity indicator affects the results of this study. In Table~\ref{tabB1} we summarise the adopted calibrations and the corresponding references. We note that although some of the calibrators are multi-valued, they present a monotonic behaviour in the covered range of abundances of this study (8.2 < $\rm 12+\log(O/H)$ < 8.6), and they can therefore be used to unambiguously determine the gas metallicity for all galaxies in the sample.

\begin{table}[h]
\caption{Alternative calibrations to derive the oxygen abundance.}
\label{tabB1}      
\centering          
\begin{tabular}{c@{\hspace{1cm}}c}   
ID & Reference \\
\hline\\[-0.355cm]\hline\\[-0.3cm]
N2-M13 & \citet{marino2013}\\[0.1cm]
N2O2-KD02 & \citet{kewley2002}\\[0.1cm]
ONS-P10 & \citet{pilyugin2010}\\[0.1cm]
ON-P10 & \citet{pilyugin2010}\\[0.1cm]
NS-P11 & \citet{pilyugin2011}\\[0.1cm]
RS32-C17 & \citet{curti2017}\\[0.1cm]
R3-C17 & \citet{curti2017}\\[0.1cm]
O3O2-C17 & \citet{curti2017}\\[0.1cm]
S2-C17 & \citet{curti2017}\\[0.1cm]
R2-C17 & \citet{curti2017}\\[0.1cm]
N2-C17 & \citet{curti2017}\\[0.1cm]
R23-C17 & \citet{curti2017}\\[0.1cm]
O3N2-C17 & \citet{curti2017}\\[0.1cm]
O3S2-C17 & \citet{curti2017}\\[0.1cm]
R23-KK04 & \citet{kobulnicky2004}\\[0.1cm]
R-P16 & \citet{pilyugin2016}\\[0.1cm]
S-P16 & \citet{pilyugin2016}\\[0.1cm]
H19 & \citet{ho2019}\\[0.1cm]
\hline                  
\end{tabular}
\end{table}

\begin{figure}
\centering
\resizebox{\hsize}{!}{\includegraphics{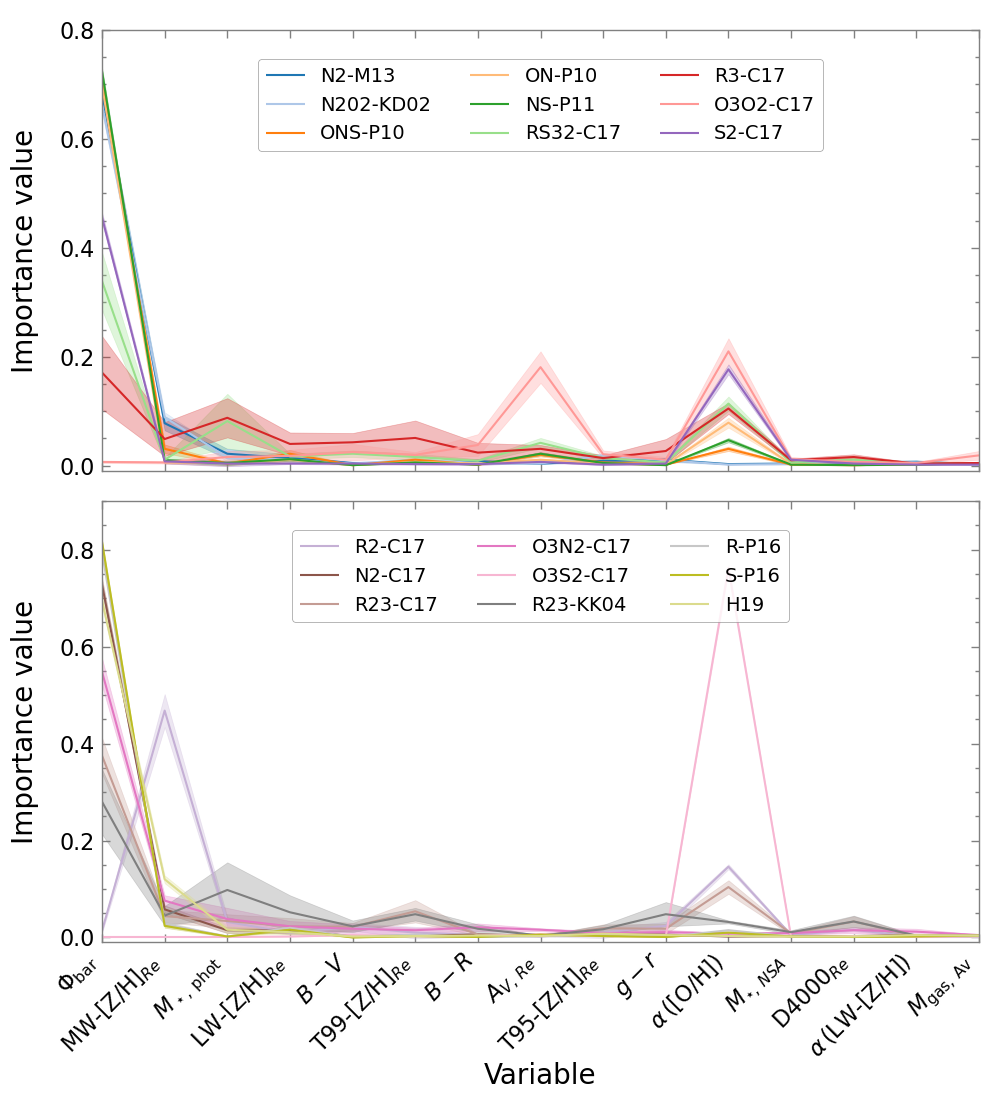}}
\caption{Importance of the 15 features of the RF with the highest importance values as shown in Figure~\ref{fig2}. The predicted $Z_g$ is estimated using different calibrations (coloured lines). See the caption of Figure~\ref{fig2} for more details.}
\label{fig1apb}
\end{figure}

In Figure~\ref{fig1apb} we show the relative importance in the RF of the 15 features with the highest importance values as ranked by the use of the main gas metallicity estimator (O3N2-M13, see Sec.~\ref{sec:results1}). Each colour corresponds to the choice of an alternative calibration (see Table~\ref{tab2} for a description of the labels). Consistent with the use of the O3N2-M13 indicator, in 15 out of the 18 cases we obtain that the baryonic gravitational potential is the best predictor of the global gas metallicity. The only exceptions are when the O3O2, the R2, and the O3S2 indicators proposed in \citet{curti2017} are used, for which the oxygen abundance gradient (two cases) or the stellar metallicity (one case) are found to be more relevant. The small dynamical range of these estimators might cause these differences. Figure~29 of \citet{sanchez2022} shows that while the other calibrators are clearly correlated with the abundances derived from H19 (arbitrarily selected as the fiducial calibrator), O3O2, R2, and O3S2 show a flatter distribution, which could bias the results and therefore dismiss their validity as abundance estimators for this work.

\begin{figure*}
\centering
\resizebox{\hsize}{!}{\includegraphics{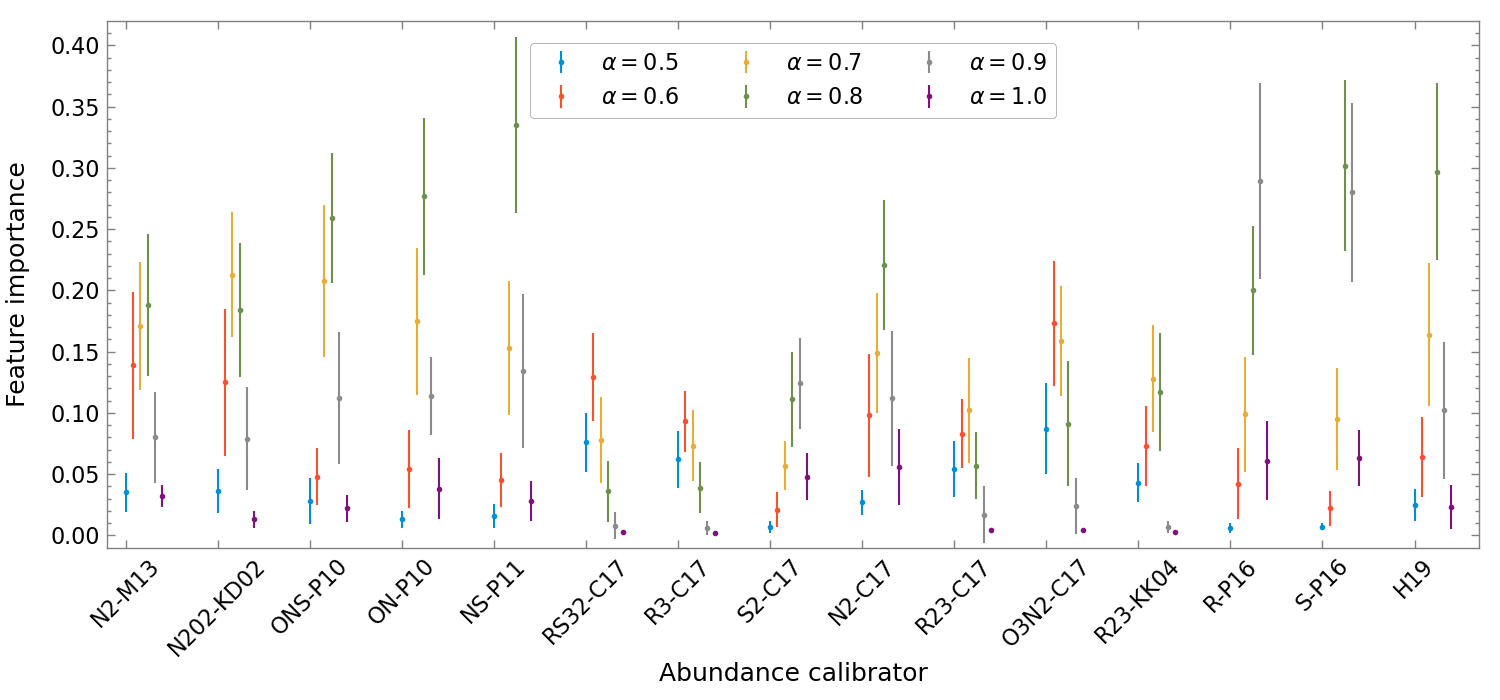}}
\caption{Importance of the six combinations of $M_\star / R_e^{\,\alpha}$ with the overall highest importance values in the RF for the alternative abundance calibrations (x-axis labels; see Table~\ref{tabB1} for references). We represent the average values (small circles) for 50 runs of the algorithm together with their standard deviations (error bars).}
\label{fig2apb}
\end{figure*}

For the 15 abundance indicators for which $\Phi_{\rm baryon}$ was found to be the most relevant parameter in the RF, we again ran the algorithm now including in the model 20 different combinations of stellar mass and galaxy size in the shape of $M_\star / R_e^{\,\alpha}$, with $\alpha=0.1,..., 2$ in steps of 0.1 (see Sec.~\ref{sec:results2} for more details). We show in Figure~\ref{fig2apb} the resulting average importances (out of 50 realisations; the error bars represent the standard deviations) of the 6 $\alpha$ values that present the highest values overall. In general, the coefficients $\alpha=0.7$ and $\alpha=0.8$ occupy the highest position in the ranking. The coefficient  $\alpha=0.6$, which was the parameter with the strongest effect on $Z_g$ when estimated with O3N2-M13, presents the highest importance value for 3 out of the 15 calibrations, but it is the third parameter in the ranking overall when the results of all abundance indicators are averaged. In Sec.~\ref{sec:results2} we showed that a scale of the form $M_\star / R_e^{\, 0.6}$ for the total gravitational potential matches the theoretical relation between the DM fraction and the baryonic surface density predicted in \citet{nestorshachar2023} very well. However, as we argued in Sec.~\ref{sec:dis}, the scatter in the relation, as well as the fact that other simulations provide different slopes for the dependence of $f_{DM}$ with $\Sigma_{\star}$, mean that the interpretation is still reasonable that $\Phi$, traced by the parameter $M_\star / R_e^{\,\alpha}$, is the best predictor of $Z_g$. We refer to that section for further discussions of the matter.

\FloatBarrier

\section{Effect of a non-homogeneous mass coverage of the sample}\label{sec:appendix3}

\begin{figure}
\centering
\resizebox{\hsize}{!}{\includegraphics{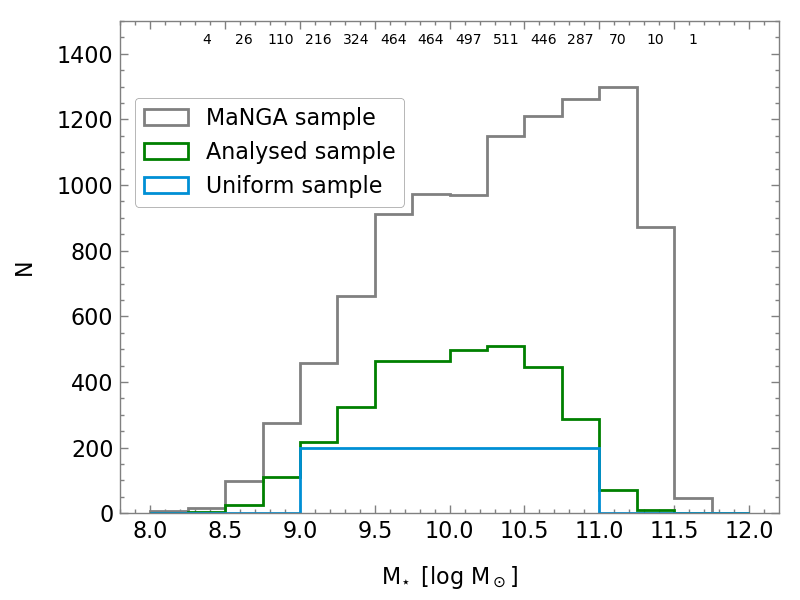}}
\caption{Stellar mass distribution of the galaxies analysed in this work (green) in comparison with the distribution of the subsample selected for this test (blue), characterised by a uniform distribution for the stellar mass. The numbers represented at the top indicate the number of galaxies from the original sample comprising each mass bin of 0.25 log(M$_\odot$). The MaNGA mother sample is also shown in grey.}
\label{fig1apc}
\end{figure}

The stellar mass distribution of the galaxies in the sample is not uniform, but there is an excess of massive systems compared to low-mass objects. This bias is inherited from the MaNGA mother sample. In order to demonstrate that the dominance of massive galaxies in the analysis does not bias the results, we selected a subsample of 1600 galaxies presenting a uniform distribution of stellar masses (i.e. a constant number of objects per mass bin). To do this, we randomly selected 200 galaxies in eight mass ranges of 0.25 $\rm \log M_{\odot}$ width from 9.0 to 11.0 $\rm \log M_{\odot}$. In Figure~\ref{fig1apc} we represent the stellar mass distribution for the three samples: the MaNGA mother sample (in grey), the analysed sample (in green), and the subsample of uniform mass distribution (in blue). Outside the range 9-11 $\rm \log M_{\odot}$, we do not have a sufficient number of galaxies to include in the test, but given the low statistics, it is unlikely that the results are driven by these objects, which are barely 6\% of the sample. 

\begin{figure}
\centering
\resizebox{\hsize}{!}{\includegraphics{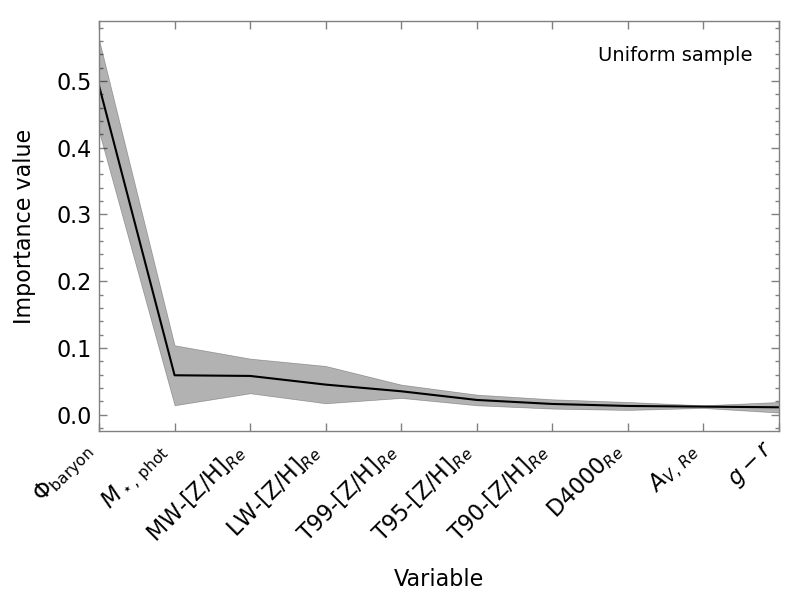}}
\caption{Average relative importance of the first ten input features in predicting the global gas metallicity for 50 runs of the RF algorithm. The features are ranked by decreasing order of importance. The shaded area represents the standard deviation of the mean relative importance of the 50 runs. The meaning of the labels is detailed in Table~\ref{tabA1}. For this test, the uniform sample was used (see Figure~\ref{fig1apc}).}
\label{fig2apc}
\end{figure}

We ran the RF using the uniform subsample, and we obtained very similar results as for the original sample. Figure~\ref{fig2apc} shows the relative importance of the first ten input features (see Table~\ref{tabA1} for a description of the labels), ranked by decreasing order of the importance value. The solid line shows the average trend of the values for 50 realisations, each of which with a different randomly selected training sample, and the shaded area represents the standard deviation of the trend. As in the original analysis, we find $\Phi_{\rm baryon}$ to be the primary factor determining the global gas metallicity, with a mean importance value of $\sim0.51$. The next parameters in the ranking are $M_\star$ and MW-[Z/H]$_{Re}$, with relative importances of $0.07$ and $0.05$, respectively. This test strengthens the main conclusion reached in this study and proves that this is not affected by any bias induced by a non-homogeneous mass coverage of the sample. All our tests show that the gravitational potential is the most important galaxy property setting the gas metallicity via the $\Phi ZR$.

\end{appendix}

\end{document}